\documentstyle[prb,aps,multicol,epsf,rotate]{revtex}

\begin{document}

\draft
\date{\today}
\title{Ferromagnetic/superconducting bilayer structure: A model system for spin
diffusion length estimation}
\author{S. Soltan $^{a,b}$ , J. Albrecht$^a$, and H.--U. Habermeier$^a$}
\address{$^a$Max--Planck--Institut f\"ur Festk\"orperforschung,
Heisenbergstr. 1, D--70569 Stuttgart, Germany}
\address{$^b$1.Physikalisches Institut, Universit\"at Stuttgart, Pfaffenwaldring 57, D--70550 Stuttgart, Germany}

\maketitle

\begin{abstract}
We report detailed studies on ferromagnet--superconductor bilayer
structures. Epitaxial bilayer structures of half metal--colossal
magnetoresistive
La$_{\mathrm{2/3}}$Ca$_{\mathrm{1/3}}$MnO$_{\mathrm{3}}$ (HM--CMR)
and high--$T_{\mathrm{c}}$ superconducting
YBa$_{\mathrm{2}}$Cu$_{\mathrm{3}}$O$_{\mathrm{7-\delta}}$(HTSC)
are grown on SrTiO$_3$ (100) single--crystalline substrates using
pulsed laser deposition. Magnetization $M$(T) measurements show
the coexistence of ferromagnetism and superconductivity in these
structures at low temperatures. Using the HM--CMR layer as an
electrode for spin polarized electrons, we discuss the role of
spin polarized self injection into the HTSC layer.  The
experimental results are in good agreement with a presented
theoretical estimation, where the spin diffusion length
$\xi_{\mathrm {FM}}$ is found to be in the range of
$\xi_{\mathrm{FM}} \approx$ 10~nm.
\end{abstract}

%

\begin{multicols}{2}
\narrowtext

\section{Introduction}
The antagonism between ferromagnetism and superconductivity has
been described very early by Ginzburg \cite{Ginzburg}. Based on
the inverse of the Meissner effect, surface currents shield the
external region from being frozen in a magnetic field. Pioneering
experiments on classical FM/SC tunneling junctions were carried
out by Tedrow and Meservey \cite{Tedrow}. In their experiment a
ferromagnetic (FM) electrode is used to inject spin--polarized
quasiparticles (SPQP) into superconducting layers. The injection
of quasiparticles into superconductors creates a local
nonequilibrium state which suppresses the superconducting order
parameter and the critical current density $j_{\mathrm{c}}$
\cite{Gray}.

Theoretical works on bilayers of metallic ferromagnets and
low-temperature superconductors \cite{Buzdin,Radovic} predicted
oscillations of the critical temperature $T_{\mathrm{c}}$. This
was confirmed by experimental results \cite{Wong,Jiang}. The
oscillating behavior of the superconducting temperature is due to
tunneling of Cooper--pairs into the FM layer \cite{Proshin}. A
review on this topic is given by Izyumov et al. \cite{Izyumov}.

In the past few years, much attention has been paid to junctions
consisting of
La$_{\mathrm{2/3}}$Ca$_{\mathrm{1/3}}$MnO$_{\mathrm{3}}$ (LCMO), a
material that shows a colossal magnetoresistance (CMR) effect, and
of YBa$_{\mathrm{2}}$Cu$_{\mathrm{3}}$O$_{\mathrm{7-\delta}}$
(YBCO), a high-$T_{\mathrm{c}}$ superconductor
\cite{Habermeier,Goldman,Yeh,Sefrioui,Holden,Albrecht,Habermeier1,Dressel}.

 Experiments with these
junctions allow to obtain information about the spin--dependent
properties of high-$T_{\mathrm{c}}$ superconductors that can lead
to the design of new superconducting devices such as ``spintronic
devices'', like transistors with high gain current and high speed.
``Spintronics'' means the exploitation of the spins of the
electrons rather than their charge. Spin controlled solid state
devices based on the giant magnetoresistance (GMR) effect are
already realized in read-out heads of hard disks. Further
challenges in the field of spintronics that are addressed by
experiments and theory include the optimization of electron--spin
life times and the detection of the spin coherence length in
nanoscale structures.

The high spin polarization, in addition to ferromagnetic ordering,
is characteristic for hole--doped rare earth manganites of the
form R$_{\mathrm{1-x}}$A$_{\mathrm{x}}$MnO$_{\mathrm{3}}$ ($R$:
trivalent rare--earth ions, $A$: divalent alkaline--earth ions).
In this case the spin polarization of the transport electrons is
close to 100~\% \cite{Soulen,Osofsky,Osofsky2}.

The fact that the in--plane lattice parameters of
La$_{\mathrm{2/3}}$Ca$_{\mathrm{1/3}}$MnO$_{\mathrm{3}}$ and
YBa$_{\mathrm{2}}$Cu$_{\mathrm{3}}$O$_{\mathrm{7-\delta}}$ are
very similar allows an epitaxial growth of LCMO/YBCO bilayers with
structurally sharp interfaces. These bilayers represent adequate
model systems to investigate spin diffusion into high-temperature
superconducting films which is shown in this paper.

 Recently, several experimental and theoretical efforts have been
performed\cite{Yeh,Wei,Gim} to determine the spin diffusion length
$\xi_{\mathrm{FM}}$ in CMR/HTSC multilayered structures, but up to
now no clear answer could be given. We show, that in bilayer or
superlattice structures of manganites and cuprates the spin
diffusion length can be determined from the experimentally
observed properties.

We present a rough theoretical estimation  for the spin diffusion
length in heterostructures of cuprates and manganites. This
estimation is able to describe properly the experimental results
of bilayer structures and superlattices of CMR/HTSC.

\begin{figure}[h!]
\epsfxsize=.3\textwidth
\centerline{{\epsffile{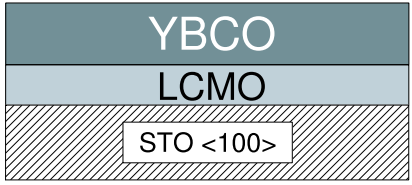}}}\vspace{0.3cm}
 \caption{Sketch of the chosen sample geometry. The
La$_{\mathrm{2/3}}$Ca$_{\mathrm{1/3}}$MnO$_{\mathrm{3}}$ (LCMO)
and YBa$_{\mathrm{2}}$Cu$_{\mathrm{3}}$O$_{\mathrm{7-\delta}}$
(YBCO) are grown by pulsed laser deposition onto SrTiO$_3$ (100)
 single crystalline substrates.}\label{sketch}
\end{figure}

Our results are obtained by investigating epitaxial bilayer
structures of thin films of ferromagnetic LCMO and
high-$T_{\mathrm{c}}$ superconducting YBCO. A sketch of these
samples is shown in Fig. (\ref{sketch}). The temperature-dependent
magnetization of these samples after zero-field cooling to T = 5~K
has been determined to show if a coexistence of ferromagnetism and
superconductivity at low temperatures occurs. In these experiments
it is found that the superconducting transition temperature of the
YBCO film decreases strongly for thicknesses of the superconductor
of $d_{\mathrm{s}}$ = 30~nm and below. This is in contrast to YBCO
single layers, where a thickness dependence of $T_{\mathrm{c}}$
occurs only  for films with thicknesses of  well below 10~nm
\cite{Tang,Triscone}. The drop of the transition temperature for
LCMO/YBCO bilayers can now be described by applying a rough
theoretical estimation taking into account the diffusion of
spin--polarized quasiparticles with a penetration depth of the
order of 10~nm.

\section{Experimental details}
Epitaxial bilayers of
La$_{\mathrm{2/3}}$Ca$_{\mathrm{1/3}}$MnO$_{\mathrm{3}}$ and
optimally doped
YBa$_{\mathrm{2}}$Cu$_{\mathrm{3}}$O$_{\mathrm{7-\delta}}$ are
grown by pulsed laser deposition. The target used  for the
ferromagnetic layer, that is first deposited, has a nominal
composition of
La$_{\mathrm{2/3}}$Ca$_{\mathrm{1/3}}$MnO$_{\mathrm{3}}$, the
superconducting layer on top is grown by using a target
YBa$_{\mathrm{2}}$Cu$_{\mathrm{3}}$O$_{\mathrm{6.95}}$. As
substrates 5$\times$5~mm$^2$ SrTiO$_3$ (STO) (100) single crystals
are used. The substrate is kept at a constant temperature of
780~$^\circ$C, the temperature is adjusted by a far-infrared
pyrometric temperature control. During the deposition an oxygen
pressure of 0.4~mbar in case of the LCMO layer and 0.6~mbar during
the YBCO deposition is used. Afterwards the bilayer is annealed
in-situ for 30--60 minutes at 530~$^\circ$C in an oxygen pressure
of 1.0~bar. This procedure results in films of high crystalline
quality, full oxygenation, and sharp film--substrate interfaces
\cite{Habermeier}. Structural studies are carried out by x-ray
diffraction (XRD) at room temperature, the resistance $R(T)$ is
measured with evaporated chromium gold contacts using the standard
four-probe technique. The temperature dependence of the
magnetization $M$(T) is recorded in a magnetic field parallel to
the film plane using a Quantum Design MPMS superconducting quantum
interference device (SQUID) magnetometer.
\begin{figure}[h!]
\epsfxsize=.35\textwidth
\centerline{{\epsffile{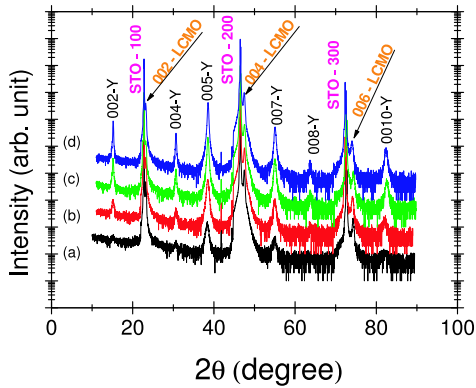}}}\vspace{0.3cm} \caption{(Color
online)~X-ray pattern  for the bilayer structures with thicknesses
of $d_{\mathrm{LCMO}}$ = 50~nm and $d_{\mathrm{YBCO}}$= 20~nm (a),
30~nm (b), 50~nm (c), and 100~nm (d) respectively. Only the
(00$\ell$) peak is found, i.e. the both layers show c--axis
textured growth on the STO (100) substrate. }\label{xrd}
\end{figure}
\section{Results and Discussion}
\subsection{X-ray diffraction pattern (XRD)}
The STO substrate has a perfect cubic perovskite structure and
both the magnetic
La$_{\mathrm{2/3}}$Ca$_{\mathrm{1/3}}$MnO$_{\mathrm{3}}$ layer and
the superconducting
YBa$_{\mathrm{2}}$Cu$_{\mathrm{3}}$O$_{\mathrm{7-\delta}}$ layer
have an orthorhombic lattice structure. The important structural
information of all of the used materials is given in table
(\ref{tab:table1}). The epitaxial YBCO thin film on top of the
LCMO layer grows under tensile strain of $\varepsilon \approx$
0.4~\%, where $\varepsilon$ can be written as: $\varepsilon$ =
$[(a_{\mathrm{LCMO}}\times
b_{\mathrm{LCMO}})^{1/2}-(a_{\mathrm{YBCO}}\times
b_{\mathrm{YBCO}})^{1/2}]/(a_{\mathrm{YBCO}}\times
b_{\mathrm{YBCO}})^{1/2}$.

\begin{minipage}{9cm}
\begin{table}[h!]
\caption{ The crystal structure data-base for the all used
materials SrTiO$_3$,
La$_{\mathrm{2/3}}$Ca$_{\mathrm{1/3}}$MnO$_{\mathrm{3}}$, and
YBa$_2$Cu$_3$O$_7$. }\label{tab:table1}
\begin{tabular}{cccccc}
 Material&\multicolumn{3}{c}{Lattice Parameter (\AA)}& Structure& Space group\\
 &a&b&c \\ \hline
SrTiO$_3$&3.905& & &cubic& Pm$\overline{3}$m(221)\\
La$_{\mathrm{2/3}}$Ca$_{\mathrm{1/3}}$MnO$_{\mathrm{3}}$&3.868 &3.858&5.453&orthorhombic&Pbnm(62)\\
YBa$_2$Cu$_3$O$_7$&3.817 &3.883 &11.682&orthorhombic&Pmmm(47)\\
\end{tabular}
\end{table}
\end{minipage}
The x-ray diffraction pattern, Fig. (\ref{xrd}a-d), shows
(00$\ell$) diffraction peaks for both the LCMO and the YBCO layers
for bilayer dimensions of $d_{\mathrm{LCMO}}$= 50~nm /
$d_{\mathrm{YBCO}}$= 20~nm Fig. (\ref{xrd}a), $d_{\mathrm{LCMO}}$
= 50~nm / $d_{\mathrm{YBCO}}$ = 30~nm Fig. (\ref{xrd}b),
$d_{\mathrm{LCMO}}$ = 50~nm / $d_{\mathrm{YBCO}}$= 50~nm Fig.
(\ref{xrd}c), and $d_{\mathrm{LCMO}}$ = 50~nm /
$d_{\mathrm{YBCO}}$= 100~nm Fig. (\ref{xrd}d); i.e. they are at
least c-axis textured. Furthermore, the sequence of the layers has
been changed. YBCO was first grown as bottom layer (here the YBCO
structure is controlled by STO substrate) and the LCMO is grown on
top. For these samples the same results are found.

 In transmission
electron microscopy (TEM) images collected at similar samples just
consisting of more than two layers \cite{Habermeier} (prepared
under the same conditions in the identical set-up) it can be found
that these samples show atomic flat interfaces. That means, that
LCMO and YBCO grow cube on cube by forming structurally a
high--quality interface. Consequently we assume that possible
perturbations created due to structural variations at the
interface can be neglected.
\begin{figure}[h!]
\epsfxsize=.35\textwidth
\centerline{{\epsffile{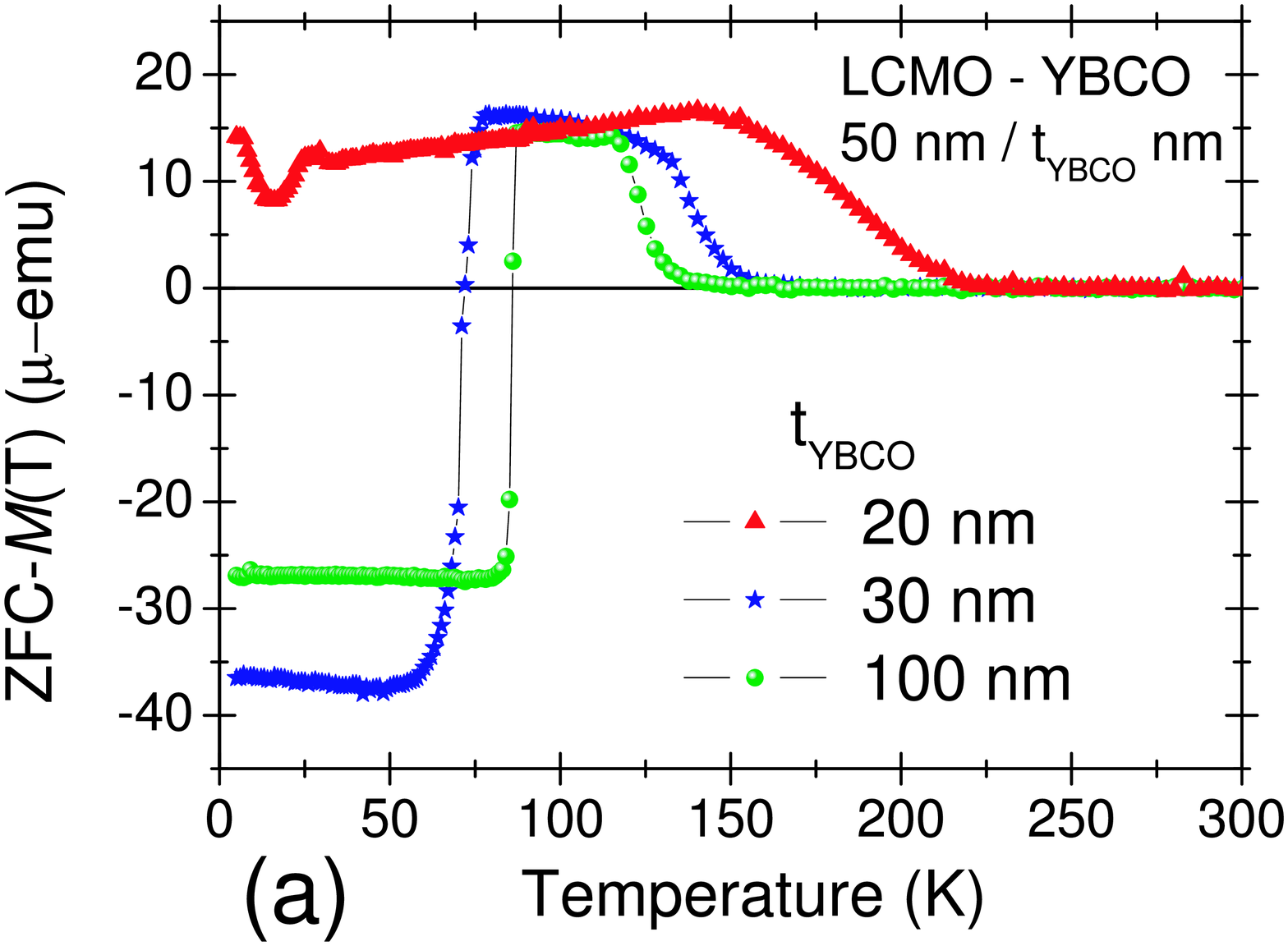}}}\vspace{0.3cm}
\epsfxsize=.35\textwidth
\centerline{{\epsffile{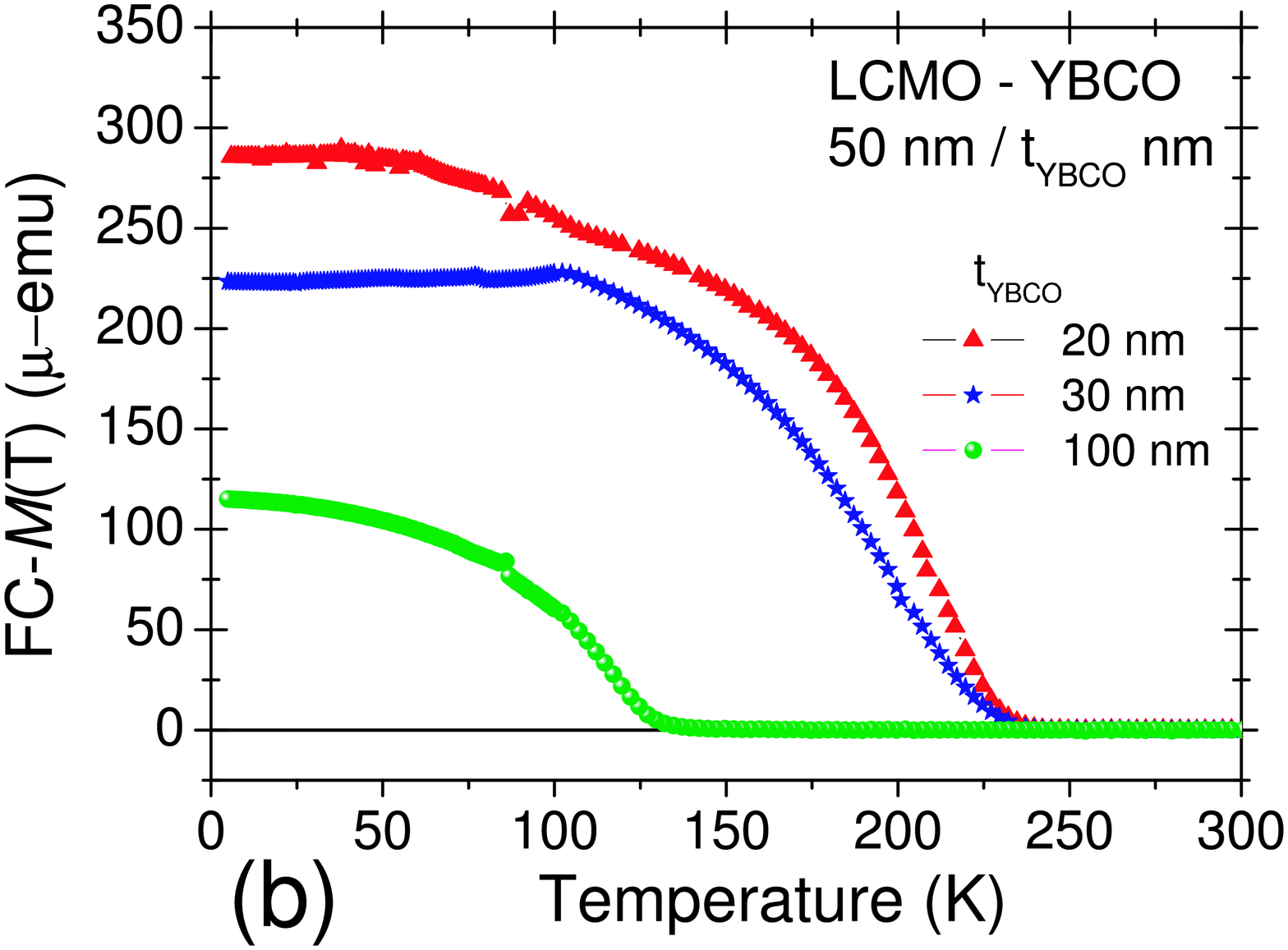}}}\vspace{0.3cm}
 \caption{(Color online)~Temperature dependence of the magnetization
 $M$(T) after zero--field cooling (a) and field--cooling (b)
   in an in--plane magnetic field $H_{\mathrm{ext}}$ = 10~Oe. Shown
   are the results
for bilayers of $d_{\mathrm{LCMO}}$ = 50~nm and
$d_{\mathrm{YBCO}}$= 20~nm, 30~nm, and 100~nm. The 50~nm / 30~nm
bilayer shows two ferromagnetic transitions in the
zero--field--cooled measurement (a). The first at
$T_{\mathrm{Curie}}$ = 229~K can not be seen in the figure the
second one occurs at $T_{\mathrm{Curie}}$ =180~K. We attribute
this behavior to a nonhomogeneous magnetic layer in that sample.
}\label{ZFC-FC}
\end{figure}
\subsection{Magnetization measurements}
Fig. (\ref{ZFC-FC}a) shows the  magnetization $M$(T) as a function
of temperature for three different bilayers. The measurements are
performed after zero--field cooling to T~=~5~K. Then, the
magnetization is measured with increasing temperature in an
in-plane external magnetic field of $H_{\mathrm{ext}}$~=~10~Oe.
The three curves refer to three different bilayers with dimensions
of $d_{\mathrm{LCMO}}$~=~50~nm and $d_{\mathrm{YBCO}}$~=~20, 30,
and 100~nm, respectively. In Fig. (\ref{ZFC-FC}a) starting at low
temperatures, we see a negative magnetization that refers to the
diamagnetic signal of the dominant superconducting state. With
increasing the temperature the magnetization jumps to positive
values. This temperature identifies the superconducting transition
temperature $T_{\mathrm{c}}$. Above the critical temperature
$T_{\mathrm{c}}$, which depends on the thickness of the YBCO
layer, we find a positive magnetization that is caused by the
ferromagnetic ordering in the LCMO layer. With increasing
temperature the magnetization drops to zero. This temperature
identifies the ferromagnetic transition temperature
$T_{\mathrm{Curie}}$. To prove that the ferromagnetic ordering is
also present below the superconducting transition, where the
signal is governed by the diamagnetic response of the YBCO layer,
also the field-cooled magnetization is measured. The results are
given in Fig. (\ref{ZFC-FC}b). It shows the magnetization $M$(T)
as function of the temperatures for the three different bilayers.
The field-cooled measurement $M$(T) is done in an in-plane
external magnetic field of $H_{\mathrm{ext}}$ = 10~Oe. Here, two
important features are found, first, ferromagnetism occurs in the
whole temperature range; $T \leq T_{\mathrm{Curie}}$. Second, the
ferromagnetic ordering shifts to lower temperatures with
increasing thickness of the YBCO layer.
\begin{figure}[h!]
\epsfxsize=.35\textwidth
\centerline{{\epsffile{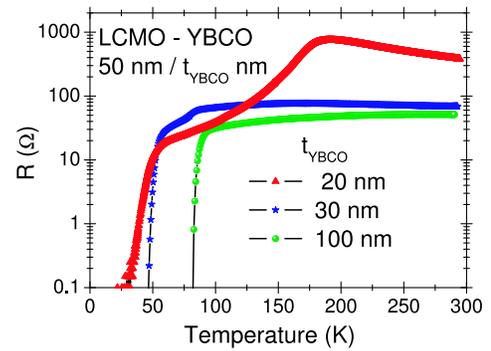}}}\vspace{0.3cm}
 \caption{(Color online)~Temperature dependent in-plane resistance $R$(T) for
 the same samples as in Fig (\ref{ZFC-FC}a). The data
  are collected from standard four-probe measurements. }\label{RT}
\end{figure}
In addition to the magnetization data we performed electric
transport measurements $R$(T) for the bilayers, as shown in Fig.
(\ref{RT}). Three results are given for $d_{\mathrm{LCMO}}$ =
50~nm and $d_{\mathrm{YBCO}}$= 20~nm, 30~nm, and 100~nm. In case
of the thinnest bilayer we find a transition from a
semiconductor--like behavior ($dR/dT$ $<$ 0) to a metallic--like
($dR/dT$ $>$ 0) around T = 180~K. This shows that the properties
of the YBCO in the bilayer have to be strongly affected due to the
presence of the LCMO layer. Otherwise the large difference in
resistivity between YBCO [$\rho_{\mathrm{YBCO}}$ (T=300~K)
$\approx 400~\mu\Omega cm$] and LCMO [$\rho_{\mathrm{LCMO}}$
(T=300~K) $ \approx 100~m\Omega cm$] would lead to current flow
only in the cuprate layer. But this behavior can only be found in
case of bilayers with thicknesses of the YBCO layer of
$d_{\mathrm{YBCO}}$= 30~nm and 100~nm, respectively. The different
$R$(T) behavior of the 50~nm/20~nm bilayer is therefore related to
a diffusion of spin--polarized quasiparticles from the LCMO layer
into YBCO.
\begin{figure}[h!]
\epsfxsize=.35\textwidth
\centerline{{\epsffile{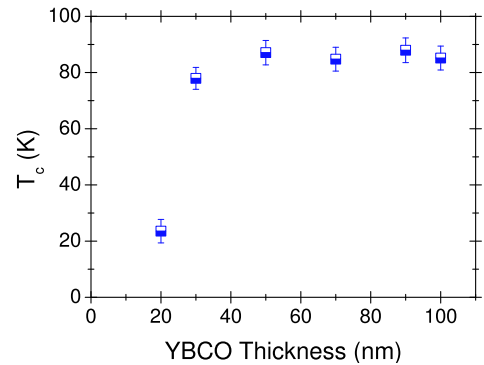}}}\vspace{0.3cm}
 \caption{(Color online)~The superconducting transition temperature of bilayers on STO with
  varying thickness of the YBCO layer obtained from the diamagnetic
   on--set in the zero--field--cooling magnetization measurements.}\label{sc}
\end{figure}
From the zero--field cooled magnetization curves $M$(T) we extract
the transition temperature $T_{\mathrm{c}}$ of the YBCO film. Here
the diamagnetic signal is used to define $T_{\mathrm{c}}$, the
values are determined by the maximum of the first derivative of
$M$(T). As a result we find a strong decrease of $T_{\mathrm{c}}$
for bilayers containing thin YBCO layers. In case of 20~nm YBCO
film we find $T_{\mathrm{c}}$= 23~K, for the 30~nm film we find
$T_{\mathrm{c}}$= 76~K, whereas bilayers with thicker YBCO films
show transition temperatures between $T_{\mathrm{c}}$= 85~K and
90~K. A plot of the transition temperatures for different YBCO
thicknesses shown in Fig.(\ref{sc}).

Additionally, we want to remark, that not only the transition
temperature but also the critical current density in the YBCO
layer is affected by the magnetic layer. Magnetic flux pinning
inside the superconductor leads to a hysteretic behavior of the
critical current density $j_{\mathrm{c}}$
\cite{Albrecht,Habermeier1}.
\begin{figure}[h!]
\epsfxsize=.35\textwidth
\centerline{{\epsffile{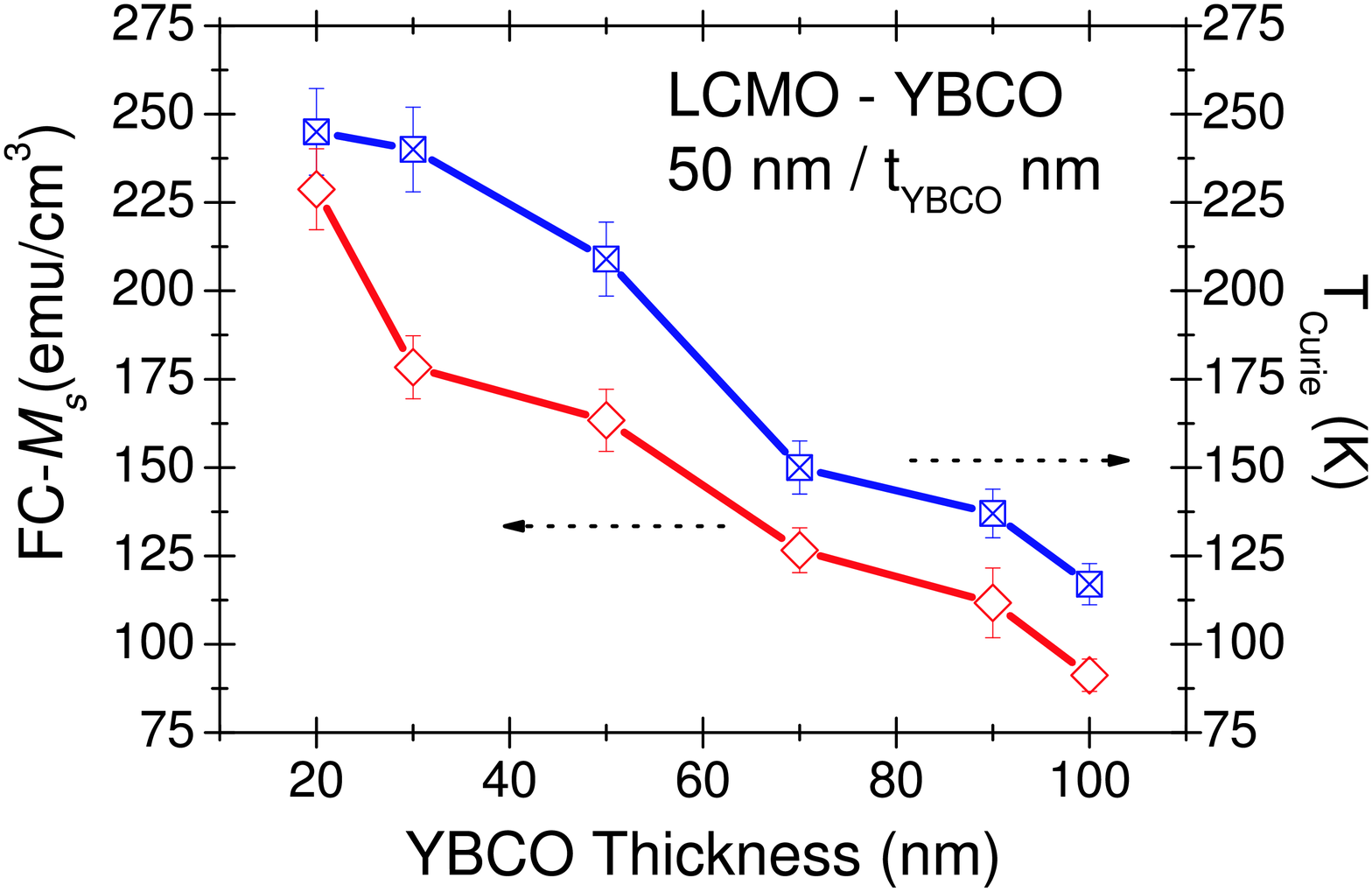}}}\vspace{0.3cm}
 \caption{(Color online)~Ferromagnetic ordering temperature $T_{\mathrm{Curie}}$
and saturation magnetization $M_s$ of bilayers with varying
thickness of the YBCO layer. $T_{\mathrm{Curie}}$ is determined
from the on--set of the ferromagnetic signal in
Fig.(\ref{ZFC-FC}b). $M_s$ is determined from the saturation
 signal $M(T)$ in Fig. (\ref{ZFC-FC}b) divided by the volume of the
LCMO layer. $T_{\mathrm{Curie}}$ and $M_{\mathrm{s}}$ strongly
decrease with increasing YBCO thickness.
  }\label{FM}
\end{figure}
Another interesting observation in  the field--cooled
magnetization $M$(T) in Fig. (\ref{ZFC-FC}b) is the increase of
the ferromagnetic transition temperature with decreasing thickness
of the YBCO layer. We measure, for a fixed CMR layer thickness of
50~nm, ferromagnetic transition temperatures $T_{\mathrm{Curie}}$
= 245~K, 240~K, and 117~K for the YBCO thicknesses of 20, 30, and
100~nm, respectively. In Fig. (\ref{FM}) a plot of the
ferromagnetic ordering temperature $T_{\mathrm{Curie}}$ of all
bilayer structures as a function of the thickness of the YBCO
layer is shown. The 1/3-Ca doped
La$_{\mathrm{1-x}}$Ca$_{\mathrm{x}}$MnO$_{\mathrm{3}}$ compound
has a bulk value of $T_{\mathrm{Curie-bulk}}$ $\approx$ 275~K, for
thin films $T_{\mathrm{Curie}}$ = 245~K is measured \cite{Praus}.
The reduction of the ferromagnetic ordering temperature may be
regarded as indication for a charge transfer from the
ferromagnetic layer into the superconductor. However, evidence for
this process can not be given from these experiments. We want to
address this question by the performance of optical conductivity
measurements. These results will be published elsewhere.
Concerning the coexistence of the two phenomena (ferromagnetism
and superconductivity), we speculate that, the very low in--plane
coherence length $\xi_{\mathrm{ab}}$ $\approx$ 1.6~nm of the
high-$T_{\mathrm{c}}$ materials rules out that such a coexistence
can be found in more than 1 to 2 unit cells away from the
interface.

\subsection{ Rough theoretical estimation of $\xi_{\mathrm{FM}}$ }
In this section we present a rough theoretical estimation that is
able to describe the observed reduction of the superconducting
transition temperature in the investigated bilayer systems. The
application of the model allows us finally to give an estimate for
the spin diffusion length of the spin polarized quasiparticles
from the ferromagnet into the superconductor.

The decay length of the superconducting order parameter
$\xi_{\mathrm{prox}}$ in the ferromagnetic layer
(\textit{proximity effect}) can be written as
\begin{equation}
 \xi_{prox} = \frac{\hbar
v_{F}}{ \Delta  E_{ex}}\label{subeq:1}
\end{equation}
where $v_{\mathrm{F}}$ is the Fermi velocity and $\Delta
E_{\mathrm{ex}}$ the exchange splitting energy of the
ferromagnetic layer.

So far, oscillations of the superconducting transition temperature
$T_{\mathrm{c}}$ in the CMR / \textit{d}-wave--HTSC superlattices
have not been found experimentally \cite{Habermeier,Sefrioui}.
This can be understood from the following equation:
\begin{equation}
 \xi_{sc} =
 \sqrt{\frac{\hbar D_{sc}}{k_B T_c}}
\label{subeq:2}
\end{equation}
where $\xi_{\mathrm{sc}}$ is the coherence length;
$D_{\mathrm{sc}}$ the electron diffusion coefficient in the
superconductor; $k_{\mathrm{B}}$  is Boltzmann's constant and
$T_{\mathrm{c}}$ the critical temperature.

From equations (\ref{subeq:1}) and (\ref{subeq:2}) we conclude
that oscillations of the critical temperature $T_{\mathrm{c}}$ of
the unconventional superlattices (CMR/\textit{d}-wave SC) do not
occur due to the large exchange energy $\Delta
E_{\mathrm{ex}}\approx$ 3~eV \cite{Quijada} for the hole-doped
rare--earth manganites in conjunction with the short coherence
length $\xi_{\mathrm{ab}}~\approx$~1.6~nm,
$\xi_{\mathrm{c}}\approx$~0.3~nm for YBCO. Additionally, a very
small spin diffusion length $\xi_{\mathrm{FM}}$ into the
superconducting layer is expected. The nearly full spin
polarization (spin$\uparrow$ or spin$\downarrow$) at the Fermi
level of LCMO leads to quenching not only  of the \textit{Andreev
reflections}\cite{Andreev} but also of the \textit{proximity
effect}, since it  prevents the Cooper--pairs to tunnel into the
magnetic layer. This also leads to the absence of oscillations of
$T_{\mathrm{c}}$ in these CMR/HTSC superlattices.

The pair breaking in CMR/HTSC due to the injection of
quasiparticles (QPI) into the superconducting layer has been taken
into account\cite{Gray,Yeh,Sefrioui}. This phenomenon has been
very early investigated by Parker \cite{Parker} and can be written
as:

\begin{equation}
 \frac{\Delta (n_{qp})}{\Delta(0)}
 \approx 1-
\frac{2n_{qp}}{4N(0)\Delta(0)} \label{subeq:3}
\end{equation}
where $\Delta (n_{\mathrm{qp}})$ is the energy required to
suppress the order parameter of the superconductor due to the
density of spin polarized quasiparticles $n_{\mathrm{qp}}$. N(0)
and $\Delta$(0) give the density of states and the order parameter
at $T$ = 0~K, respectively. $n_{\mathrm{qp}}$ is generated by
self-injection along the c-axis across the highly transparent
interface and is governed by the high exchange splitting energy
$\Delta E_{\mathrm{ex}}~\approx$~3~eV of the magnetic layer, see
equation (\ref{subeq:1}). The QPI is a temperature dependent
function and can be derived in the following form
\cite{Gim,Nicol}:

\begin{equation}
 n_{qp}(T) \approx 4N(0)\Delta(0)
\sqrt{\frac{\pi}{2}\frac{\Delta(T)k_BT}{\Delta^2(0)}}~~
e^\frac{-\Delta(T)}{k_{B}T} \label{subeq:4}
\end{equation}
The spin diffusion length $\xi_{\mathrm{FM}}$ can now be
determined after Ref.\cite{Gray} analogous to a classical FM/SC
structure as:

\begin{equation}
 \xi_{FM} \approx \sqrt{\ell_o
v_F \tau_s} \label{subeq:5}
\end{equation}

Here $\ell_o$(T = 0~K) $\approx$ 20~nm is the mean free path in
YBCO \cite{Yeh}, the spin diffusion relaxation time $\tau_s$ is
given by:

\begin{equation}
 \tau_s \approx 3.7 \frac{\hbar k_B
T_c}{ \Delta E_{ex}
 \Delta(T)}\label{subeq:6}
\end{equation}

where:
\begin{equation}
\Delta(T)\sim \Delta(0) \sqrt{1- (\frac{T}{T_c})} \label{subeq:7}
\end{equation}

with $\Delta(0)~\approx$~20~meV for YBCO \cite{Yeh}. From equation
(\ref{subeq:3}), we end up with a relation where the temperature
dependence and the length scale of the spin diffusion length
$\xi_{\mathrm{FM}}$ is included.

First, we have to consider the spin density in the superconductor.
It is assumed that spins in high--temperature superconductors can
be described as unitary scatterers. From Zn doping in YBCO it is
known that a critical doping in the range of 2-10~\% strongly
reduces $T_{\mathrm{c}}$ \cite{Hirschfeld,Franz,Yang}. This
critical density of spins is achieved at a distance of $d$ =
$\alpha$
 $\xi_{\mathrm{FM}}$, with $\alpha~\approx$~3. This $d$ is now
identified with the YBCO film thickness. This enables us to model
the experimental data by:
\begin{equation}
d = \alpha\xi_{FM} \cong 3.7 \frac{\alpha m^* \hbar
v_F^2}{\Delta(0) \Delta E_{ex} n_{qp}(0)e^2}
  \frac{\sqrt{T/n_{qp}(T)}}{\sqrt[4]{1-(T/T_c)}}
 \label{subeq:8}
\end{equation}
$m^\ast$ and $e$ are the electron effective mass and charge.

\begin{figure}[h!]
\epsfxsize=.35\textwidth
\centerline{{\epsffile{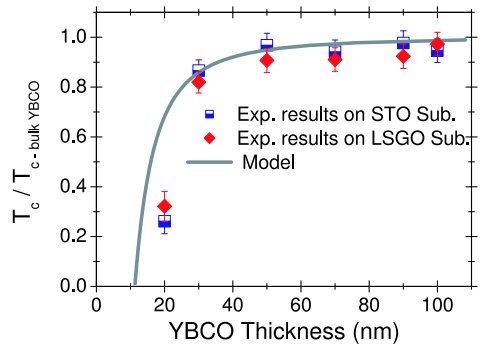}}}\vspace{0.3cm}
 \caption{ (Color online)~The normalized superconducting transition temperature of bilayers with
  varying thickness of the YBCO layer on STO and LSGO substrates obtained from the diamagnetic
   on--set in the zero--field--cooling magnetization measurement (squares) and
   model according to equation (\ref{subeq:8}) (solid line).
   The model gives a good description of the experimental data.}\label{Model}
\end{figure}

Introducing now $n_{\mathrm{qp}}$(T) from equation (\ref{subeq:4})
we are able to fit our experimental data using the quasiparticle
density as the only free parameter. This parameter
$n_{\mathrm{qp}}$(T) describes the decrease of the energy gap
$\Delta$($n_{\mathrm{qp}}$)\cite{Parker,Owen}. In our case we find
the best fitting for $n_{\mathrm{qp}}$(T)$\approx$ 0.36, 0.35, and
0.13 for T/$T_{\mathrm{c}}$~=~0.01, 0.3, and 0.9, respectively,
which agrees with other theoretical calculations of
$n_{\mathrm{qp}}$(T)$\approx$~0.32, 0.31, and 0.09 at the same
temperature ratio for \textit{d}-wave superconductors
\cite{Nicol,Koller}. Figure (\ref{Model}) shows the experimentally
determined transition temperatures normalized to the
$T_{\mathrm{c-bulk}}$=91~K for different film thicknesses on two
different substrates SrTiO$_{\mathrm{3}}$ (STO) and
LaSrGaO$_{\mathrm{4}}$ (LSGO) and the fit to the rough theoretical
estimation.
The description fits nicely to the experimental results and
suggests that the recovery of the transition temperature
$T_{\mathrm{c}}$ of the YBCO top layer takes place at about 30~nm
which leads to a spin diffusion length of $\xi_{FM} = d /
\alpha~\approx$10~nm. This finding is in good agrement with
results that have been estimated by Holden et al. \cite{Holden}
from  optically investigated LCMO/YBCO superlattices by
spectroscopic ellipsometry measurements of the far--infrared (FIR)
dielectric properties. These results provide an evidence that the
free carrier response is strongly suppressed in these
superlattices as compared to pure YBCO and LCMO films, and they
estimate that a critical thickness for the YBCO is in the range of
20~nm. Note, that in case of superlattices the spin diffusion
quasiparticles penetrate from both sides into to the
superconducting film. The accordance between the results shown in
this paper and other groups using different experimental
techniques  gives rise to a spin diffusion length
$\xi_{\mathrm{FM}}$ from LCMO into YBCO in the order of 10~nm at
low temperatures.
\section{conclusion}
We have investigated experimentally and  estimated theoretically
the effects of the diffusion of spin polarized quasiparticles in
bilayer structures of  manganites and  cuprates. From x-ray
measurements we have shown that
YBa$_{\mathrm{2}}$Cu$_{\mathrm{3}}$O$_{\mathrm{7-\delta}}$ thin
films can be grown epitaxially on thin epitaxial films of
La$_{\mathrm{2/3}}$Ca$_{\mathrm{1/3}}$MnO$_{\mathrm{3}}$.
Transport and magnetization measurements show the coexistence of
ferromagnetism and superconductivity  at low temperatures in these
structures. We find that the transition temperature of the
superconducting film drastically decreases with thinner
YBa$_{\mathrm{2}}$Cu$_{\mathrm{3}}$O$_{\mathrm{7-\delta}}$ films.
The development of a simple model allows us to explain the
experimental data and enables us to determine the spin diffusion
length of spin polarized quasiparticles from
La$_{\mathrm{2/3}}$Ca$_{\mathrm{1/3}}$MnO$_{\mathrm{3}}$ into
YBa$_{\mathrm{2}}$Cu$_{\mathrm{3}}$O$_{\mathrm{7-\delta}}$ to be
in the range of $\xi_{\mathrm{FM}} \approx$ 10~nm.
\begin{acknowledgments}
The authors are grateful to G. Cristiani for the preparation of
the outstanding samples, and E. Br\"ucher for SQUID measurements.
S.S. is grateful to Max--Planck--Society and Ministry of Higher
Education and Scientific Research, Egyptian government for
support.
\end{acknowledgments}

 {

 }

\end{multicols}

\references




\bibitem{Ginzburg} V.L. Ginzburg, Sov. Phys. JETP \textbf{4},
153 (1957).


\bibitem{Tedrow} P.M. Tedrow and R. Meservey, Phys. Rev. Lett. \textbf{26}, 192 (1971).


\bibitem{Gray} K.E. Gray, in ``Nonequilibrium Superconductivity, Phonons and Kapitza
Boundaries'' , edited by K. E. Gray (Plenum, New York, 1981).



\bibitem{Buzdin} A.I. Buzdin and M. Y. Kupriyanov, JETP Letters \textbf{52}, 487
(1990).



\bibitem{Radovic} Z. Radovic, M. Ledvij, L. Dobrosavljevic´, A.I. Buzdin, and J.R.
Clem, Phys. Rev. B \textbf{44}, 759 (1991), and references
therein.



\bibitem{Wong} H.K. Wong, B. Y. Jin, H.Q. Yang, J.B. Ketterson, and J.E.
Hilliard, J. Low Temp. Phys. \textbf{63}, 307 (1986).



\bibitem{Jiang} J.S. Jiang, D. Davidovic´, D.H. Reich, and C.L. Chien, Phys.
Rev. Lett.\textbf{74}, 314 (1995), and Rev. B \textbf{54}, 6119
(1996).



\bibitem{Proshin} Yu. N. Proshin and M.G. Khusainov, JETP \textbf{86}, 930 (1998)



\bibitem{Izyumov} Yu.A.  Izyumov, Yu.N.  Proshin, and M.G. Khusainov, Phys. Usp.
\textbf{45}, 109 (2002), and references therein.



\bibitem{Habermeier} H.-U. Habermeier, G. Cristiani, R.K. Kremer,
O. Lebedev, G. Van Tendeloo, Physica C \textbf{364}, 298 (2001).



\bibitem{Goldman} A.M. Goldman, V.A. Vas'ko, P.A. Kraus, K.R. Nikolaev,
and V.A. Larkin, J. Magn. Magn. Mater. \textbf{200}, 69 (1999).





\bibitem{Yeh} N.C. Yeh, R.P. Vasquez, C.C. Fu, A.V. Samoilov, Y. Li, and K. Vakili,
Phys. Rev. B \textbf{60}, 10522 (1999).



\bibitem{Sefrioui} Z. Sefrioui, D. Arias, V. Pen˜a, J.E. Villegas, M. Varela,
P. Prieto, C. Leo´n, J.L. Martinez, and J. Santamaria, Phys. Rev.
B \textbf{67}, 214511 (2003).



\bibitem{Holden} T. Holden, H.-U. Habermeier, G. Cristiani, A. Golnik, A. Boris, A. Pimenov,
J. Huml\'{i}\v{c}ek,  O. Lebedev, G. Van Tendeloo, B. Keimer, and
C. Bernhard,  Phys. Rev. B \textbf{69}, 064505 (2004).




\bibitem{Albrecht} J. Albrecht, S. Soltan,  and H.-U. Habermeier, EuroPhys.
Lett.  \textbf{63}, 881 (2003).



\bibitem{Habermeier1} H.-U. Habermeier, J. Albrecht, and S. Soltan,
Supercond. Sci. Technol. \textbf{17}, S140 (2004).




\bibitem{Dressel} F. Chen, B. Gorshunov, G. Cristiani, H.-U. Habermeier, and M.
Dressel, Solid State Commun. \textbf{131}, 295 (2004).



\bibitem{Soulen} R.J. Soulen, J.M. Byers, M.S. Osofsky, B. Nadgorny, T.
Ambrose, S.F. Cheng, P. R. Broussard, C.T. Tanaka, J. Nowak, J.S.
Moodera, A. Barry, and J.M.D. Coey, Science \textbf{282}, 85
(1998).



\bibitem{Osofsky} M.S. Osofsky, R.J. Soulen, B.E. Nadgorny, G. Trotter, P.R.
Broussard, and W.J. Desisto, Mater. Sci. Eng. B \textbf{84}, 49
(2001).




\bibitem{Osofsky2} M.S. Osofsky, B. Nadgorny, R.J. Soulen, P. Broussard, and M.
Rubinstein, J. Byers, G. Laprade, Y.M. Mukovskii, D. Shulyatev,
and A. Arsenov, J. Appl. Phys.\textbf{ 85}, 5567 (1999).




\bibitem{Wei} J.Y. Wei, J. Superconduct. \textbf{15}, 67 (2002).




\bibitem{Gim} Y. Gim, A. W. Kleinsasser, and J. B. Barner, J. Appl. Phys. \textbf{90}, 4063 (2001).




\bibitem{Tang} W.H. Tang, C.Y. Ng, C.Y. Yau, and J. Gao, Supercond. Sci. Technol. \textbf{13}, 580 (2000).




\bibitem{Triscone} J.M. Triscone, $\O$. Fischer, O. Brunner, L. Antognazza, A.D. Kent, and
M.G. Karkut, Phys. Rev. Lett. \textbf{64}, 804 (1990).




\bibitem{Praus} R.B. Praus, G.M. Gross, F.S. Razavi, and  H.-U.
Habermeier, J. Magn. Magn. Mater. \textbf{211}, 41 (2000).



\bibitem{Quijada} M. Quijada, J. Cerne, J.R. Simposon, H.D. Drew,
K-H. Ahn, A.J. Millis, R. Shreekala, R. Ramesh, M. Rajeswari, and
T. Venkatesan,  Phys. Rev. B \textbf{58}, 16093 (1998).




\bibitem{Andreev} A.F. Andreev and Zh. Eksp, Sov. Phys. JETP \textbf{19}, 1228 (1964).



\bibitem{Parker} W.H. Parker,  Phys. Rev. B \textbf{12}, 3667 (1975).



\bibitem{Nicol} E.J. Nicol and J.P. Carbotte, Phys. Rev. B \textbf{67},
214506 (2003).



\bibitem{Hirschfeld} P.J. Hirschfeld and N.D. Goldenfeld, Phys.
Rev. B \textbf{48}, 4219 (1993).


\bibitem{Franz} M. Franz, C. Kallin, A.J. Berlinsky, and M.I.
Salkola, Phys. Rev. B \textbf{56}, 7882 (1997).



\bibitem{Yang} C.Y. Yang, A.R. Moodenbaugh, Y.L. Wang, Youwen Xu, S.M. Heald, D.O. Welch,  M.
Suenaga, D.A. Fischer, and J.E. Penner–Hahn, Phys. Rev. B
\textbf{42}, 2231 (1990).



\bibitem{Owen} C.S. Owen and D.J. Scalapino, Phys. Rev. Lett. \textbf{28},
1559 (1972).



\bibitem{Koller} D. Koller, M.S. Osofsky, D.B. Chrisey, J.S.
Horwitz, R.J. Soulen, R.M. Stroud, C.R. Eddy, J. Kim, R.C.Y.
Auyeung, J.M. Byers, B.F. Woodfield, G.M. Daly, T.W. Clinton, and
M. Johnson, J. Appl. Phys.~\textbf{83}, 6774 (1998).
\end{document}